# Experimentally Validated Hopping Transport Model for Energetically Disordered Organic Semiconductors

Tanvi Upreti[1], Yuming Wang[2], Huotian Zhang[2], Dorothea Scheunemann[1], Feng Gao[2], Martijn Kemerink[1,*]

[1] Complex Materials and Devices, Linköping University, Department of Physics, Chemistry and Biology (IFM), SE-581 83 Linköping, Sweden
[2] Biomolecular and Organic Electronics, Linköping University, Department of Physics, Chemistry and Biology (IFM), SE-581 83 Linköping, Sweden

* email: martijn.kemerink@liu.se

## Abstract

Charge transport in disordered organic semiconductors occurs by hopping of charge carriers between localized sites that are randomly distributed in a strongly energy dependent density of states. Extracting disorder and hopping parameters from experimental data like temperature dependent current-voltage characteristics typically relies on parametrized mobility functionals that are integrated in a drift-diffusion solver. Surprisingly, the functional based on the extended Gaussian disorder model (eGDM) has been extremely successful at this, despite it being based on the assumption of nearest neighbor hopping (nnH) on a regular lattice. We here propose a variable range hopping (VRH) model that has been integrated in a freeware drift-diffusion solver. The mobility model has been calibrated using kinetic Monte Carlo calculations and shows good agreement with the Monte Carlo calculations over the experimentally relevant part of the parameter space. The model is applied to temperature-dependent space charge limited current (SCLC) measurements of different systems. In contrast to the eGDM, the VRH model provides a consistent description of both p-type and n-type devices. We find a critical ratio of $a_{NN}/\alpha$ (mean inter-site distance / localization radius) of ~3 below which hopping to non-nearest neighbors becomes important around room temperature and the eGDM cannot be used for parameter extraction. Typical (Gaussian) disorder values in the range 45-120 meV are found, without any clear correlation with photovoltaic performance when the same active layer is used in an organic solar cell.

## Introduction

Recent years have seen a considerable improvement in the field of organic photovoltaics (OPVs) where power conversion efficiencies exceeding 16% have been reached[1]. To improve upon the performance of such devices, further understanding the underlying physics of charge transport becomes important as the (transient) mobility of the generated charge carriers is one of the key factors that decide how efficiently the charges can escape recombination and be extracted[2–6]. Similar arguments can be made for the need to understand charge transport in other organic devices, including light emitting diodes, field effect transistors and a range of mixed ionic-electronic devices. However, in the following we will focus on OPV materials.

Charge transport in these systems is commonly understood to occur via thermally activated tunneling, hopping, of charge carriers between localized sites. Due to the wide ($\gg k_B T$), typically Gaussian or exponential distribution of site energies, the mobility becomes strongly dependent on temperature *T*, charge carrier density *n* and electrostatic field *F*. Various approaches have been proposed to calculate the mobility functional $\mu(T, n, F)$ for hoping transport in disordered semiconductors. The seminal work by Bässler et al. used kinetic Monte Carlo (kMC) simulations to this end.[2] While the resulting Gaussian disorder model (GDM) rationalized experimentally observed temperature dependencies, it only considered the low-density Boltzmann limit and showed discrepancies in the field dependence that later have been attributed to spatial correlations of the site energies.[7,8] The charge density dependence of the mobility beyond the Boltzmann limit was included in the so-called extended Gaussian disorder model (eGDM).[9] Pasveer et al. introduced a parametrization scheme for the corresponding mobility functional $\mu(T, n, F)$. Although these expressions are conveniently implemented in e.g. drift-diffusion solvers and have been widely used, they have also been heavily criticized.[10–12] A major point of criticism is the assumption of a regular, cubic lattice and the use of a small and constant localization radius that enforces strict nearest neighbor hopping (nnH) and gives rise to an erroneous field dependence if the assumed conditions are not met.[11] Despite the fundamental criticism, it is to date unknown how severe, from a pragmatic, experimentalist perspective, the errors are when one applies the eGDM parametrization to extract hopping and disorder parameters from experimental data. A strongly related problem is the absence of a complementary variable range hopping (VRH) model for a Gaussian density of states that is of sufficient user friendliness to be used in data analysis. A step in this direction was made in the works of Cottaar *et al*. and Nenashev *et al* who derived analytical expressions for the functional $\mu(T, n)$ for VRH on a simple cubic lattice.[10,13]

Among the various experimental techniques that are used to characterize transient and quasi-static charge transport,[5,14–18] space charge limited current (SCLC)[19] is one of the most commonly used methods to estimate the mobility of charge carriers and the underlying hopping and disorder parameters. Although being a straightforward method from an experimental perspective, the interpretation of current density vs. voltage (jV) curves obtained from (temperature-dependent) SCLC can be rather intricate[20,21]. In a recent paper, we reviewed the most commonly employed methods and sources of error and introduced an automated software to analyze SCLC(T) jV data using, amongst others, the (nnH) models discussed above.[21] While the widely used 'slope 2 fitting' method clearly led to erroneous results, the other models, with the exception of the correlated disorder model,[8] typically led to consistent results regarding the energetic disorder of hole-only diodes. Electron transport was not addressed.

The aim of the current paper is to propose and establish a VRH model for electrons and holes that is both physical and practical and that allows a consistent extraction of parameters from steady state transport measurement like SCLC. The proposed model is calibrated against numerically exact kinetic Monte Carlo simulations of VRH and integrated in the open source analysis tool described in Ref. [21]

that can be downloaded as freeware.[22] The model is used to analyze temperature dependent SCLC curves for electron- and hole-only devices from a variety of binary and ternary OPV materials. The investigated blends include both polymer:fullerene and polymer:non-fullerene blends with power conversion efficiencies ranging from ~5% to ~14%. While for a significant number of materials the nnH model can be applied, in the meaning of giving very similar hopping parameters as the VRH model, the major advantage of the VRH model is that it provides a consistent determination of the hopping parameters of all material systems. The extracted parameters allow to predict whether nnH could be used alternatively. Especially for fullerene-containing layers, electron hopping to non-nearest neighbors cannot be ignored, leading to a significant underestimation of the energetic disorder in the LUMO level if an (e)GDM analysis would be performed.

## Theory
### Analytical variable range hopping model

In case of intrinsic organic semiconductors, it is common to assume a Gaussian energy distribution for the localized states

$$g(\varepsilon) = \frac{N_i}{\sqrt{2\pi}\sigma_{DOS}} \exp\left(\frac{-(\varepsilon-\varepsilon_0)^2}{2\sigma_{DOS}^2}\right) \quad (1)$$

Here, $\sigma_{DOS}$ is the energetic disorder, typically in the range 0.05-0.1eV, $\varepsilon_0$ is the central energy of the highest occupied molecular orbital (HOMO) or the lowest unoccupied molecular orbital (LUMO) and $N_i$ is the total site density of the randomly distributed localized states that relates to the mean inter-site distance as $a_{NN} = N_i^{-1/3}$.

The charge hopping rates from a site $i$ to $j$ are typically described by either the Miller Abrahams or the Marcus expression. Although the latter appears more physical as it accounts in lowest order for the lattice deformation that is associated with the presence of a net charge on a piece of conjugated material (polaron formation), the resulting mobilities are typically rather similar for the two rates[13]. Since the Miller-Abrahams rates require one less parameter to be known, we choose to work with these rates, for which the hopping rate between the site $i$ and $j$ is given as

$$\nu_{ij} = \nu_0' \exp\left(\frac{-2r_{ij}}{\alpha}\right) \exp\left(-\frac{\Delta E}{k_B T}\right), \quad (2)$$

where $k_B$ is the Boltzmann constant, $\Delta E = \max(0, \varepsilon_j - \varepsilon_i - \vec{F} \cdot \vec{r}_{ij})$ with $\alpha$ the localization radius, $F$ the local electrostatic field vector and $r_{ij}$ the vector from $i$ to $j$. The prefactor $\nu_0' = \nu_0 \exp(2a_{NN}/\alpha)$ makes $\nu_0'$ become the rate of downward hops to a nearest neighbor site and is introduced for consistency with mobility expressions from lattice models and that we shall fix to $1\times10^{11}$ s$^{-1}$ unless stated otherwise.[9,10,21] The prefactor $\nu_0$ is typically identified as the attempt-to-hop frequency.

Numerical solutions for the current density and charge carrier mobility in finite systems with rates given by (2) and site energies randomly drawn from (1) have been reported by various authors. Specifically, Pasveer et al. used a master equation method to generate a parametrized mobility functional for hopping on a simple cubic lattice for $a_{NN}/\alpha = 10$ as[9]

$$\mu(T, c, F) \approx \tilde{\mu}(T, c) f(T, F) \quad (3)$$

The expressions for the factors $\tilde{\mu}(T, c)$ and $f(T, F)$ are given in the SI for completeness. Below, we will analyze a series of SCLC(T) experiments using either the mobility functional (3), which we shall refer to as the eGDM, or the mobilities from the analytical model that is introduced next.

The analytical VRH model is an extension of the Mott-Martens model as reviewed by Coehoorn et al.[23], along the lines previously used by Zuo et al. for doped organic semiconductors.[24,25] The conductivity is given by a Miller Abrahams-type expression

$$\sigma = \sigma_0 \exp\left[-2\alpha R^* - \frac{\varepsilon^* - \varepsilon_F}{k_B T}\right] \quad (4)$$

where the conductivity prefactor $\sigma_0 = n\mu_0$ with $n$ the charge carrier density and $\mu_0$ a mobility prefactor. Using a percolation argument, the conductivity is assumed to be governed by characteristic hops over a distance $R^*$, going from the Fermi energy $\varepsilon_F$ to the transport energy $\varepsilon^*$. $R^*$ and $\varepsilon^*$ are related through density of states (1) as

$$B_C = \frac{4}{3}\pi(R^*)^3 \int_{\varepsilon_F}^{\varepsilon^*} g(\varepsilon)\, d\varepsilon \quad (5)$$

where $B_C$ is the critical number of bonds in the infinite percolating cluster. Maximizing (4) under the conditions set by (5) allows for variable range hopping to occur. Although this will not be pursued further here, we notice that fixing $R^* = a_{NN}$ leads to a nearest neighbor hopping model that rather accurately reproduces numerical simulations of the type performed by Pasveer et al.[9]

Because of the absence of a proper analytical theory to express the non-exponential terms in (4), i.e. the mobility prefactor, we make the ansatz:

$$\mu_0(T) = B\left(\frac{a_{NN}^2 \nu_0}{k_B T}\right)\left(\frac{k_B T}{\sigma_{DOS}}\right)^{\lambda_1}\left(\frac{\alpha}{a_{NN}}\right)^{\lambda_2} \quad (6)$$

This expression is similar to what was derived in Refs. [10,13], except for the last term on the right hand side, which has been introduced in analogy to the dimensionless energy scaling factor $(k_B T/\sigma)^{\lambda_1}$. Instead of the term $a_{NN}^2$ in Eq. 6, one could expect a term $\langle R^2 \rangle$, being the expectation value of the squared hopping distance. While this leads to a slightly improved agreement with kMC results for the simulations parametric in $T$ and $\sigma_{DOS}$, it severely deteriorates the agreement for the simulations parametric in $a_{NN}$ and $\alpha$ as shown in the SI, Figs. S1 and S2.

To include the effects of finite electric fields, we use the effective temperature concept as introduced by Marianer and Shklovskii[26] and later Baranovskii et al[27]. In their work it was proposed that the combined effect of electric field and temperature can be captured by replacing the lattice temperature $T$ with an effective temperature for the charge carrier distribution:

$$T_{eff} = \left[T^2 + \left(\gamma \frac{qF\alpha}{k_B}\right)^2\right]^{\frac{1}{2}} \quad (7)$$

where $\gamma \approx 0.67$. In the following, the mobility model Eqs. (4-7) will be referred to as the effective temperature Gaussian disorder model, ET-GDM.

Prior to applying the ET-GDM model to experimental jV curves, a 'universal' set of parameters $\lambda_1$, $\lambda_2$, $B$ and $B_c$ is to be found. To this end, we phenomenologically calibrate the analytical model to mobilities obtained from numerically exact kinetic Monte Carlo simulations as a function of concentration and $F$ parametric in $\sigma$, $T$, $a_{NN}$, and $\alpha$. Although the ET-GDM model is physically transparent and leads to equations that are easily solvable by numerical means, integration in a drift-diffusion device solver that allows fitting of experimental data on reasonable time scales is not completely trivial. Hence, to make the methodology available to others, we implemented it in the automated SCLC analysis program FitSCLC that can be freely downloaded.[21,22] The same code can be used with the eGDM mobility functional as well as with a range of other analysis schemes.

## Kinetic Monte Carlo model

Numerical kinetic Monte Carlo (kMC) simulations are an established method to obtain numerically exact solutions of the transport problem posed by a finite density of interacting particles hopping in a landscape with positional and/or energetic disorder[2,6,10,28]. Here we use a standard kMC scheme to generate reference mobilities as a function of charge carrier concentration, disorder, temperature, localization radius, site density and electric field. The program is similar to the one described in Ref. [6], except that it works on a random lattice instead of a regular lattice and that it allows for variable range hopping by considering hops to a preset number of nearest neighbor sites that was chosen sufficiently large such that further increases did no longer affect the mobility; typically 63-126 neighboring sites were included. Single particle site energies were randomly selected from a Gaussian DOS (Eq. 1). Hopping rates are calculated from Eq. 2 and used as weight factors in the otherwise random selection of the hopping event in each time step. The waiting time for each time step is calculated as $\tau = -\frac{\ln(r)}{\Sigma_\nu}$ where $r$ is a random number drawn from a homogeneous distribution between 0 and 1 and $\Sigma_\nu$ is the sum of the rates of all possible events. On-site particle-particle interaction, preventing double site occupation, is included.

## Experimental

Electron and hole transport were analyzed using temperature dependent space charge limited current measurements on single carrier devices based on a number of representative OPV materials. Specifically, the fullerene-based TQ1:PC$_{71}$BM (Poly[[2,3-bis(3-octyloxyphenyl)-5,8-quinoxalinediyl]-2,5-thiophenediyl]) : ([6,6]-Phenyl C71 butyric acid methyl ester) and the non-fullerene PBDB-T:IEICO-4F (Poly[[4,8-bis[5-(2-ethylhexyl)-2-thienyl]benzo[1,2-b:4,5-b']dithiophene-2,6-diyl]-2,5-thiophenediyl[5,7-bis(2-ethylhexyl)-4,8-dioxo-4H,8H-benzo[1,2-c:4,5-c']dithiophene-1,3-diyl]]) : (2,2'-((2Z,2'Z)-(((4,4,9,9-tetrakis(4-hexylphenyl)-4,9-dihydro-sindaceno[1,2-b:5,6-b']dithiophene-2,7-diyl)bis(4-((2-ethylhexyl)oxy)thiophene-5,2-diyl))bis(methanylylidene))bis(5,6-difluoro-3-oxo-2,3-dihydro-1H-indene-2,1-diylidene))dimalononitrile) and PM6:Y6 (Poly[(2,6-(4,8-bis(5-(2-ethylhexyl-3 fluoro)thiophen-2-yl)-benzo[1,2-b:4,5-b']dithiophene))-alt-(5,5-(1',3'-di-2-thienyl-5',7'-bis(2 ethylhexyl)benzo[1',2'-c:4',5'-c']dithiophene-4,8-dione)]) : (2,2'-((2Z,2'Z)-((12,13-bis(2-ethylhexyl)-3,9-diundecyl-12,13-dihydro-[1,2,5]thiadiazolo[3,4-e]thieno[2,"3'':4',5']thieno[2',3':4,5]pyrrolo[3,2-g]thieno[2',3':4,5]thieno[3,2-b]indole-2,10-diyl)bis(methanylylidene))bis(5,6-difluoro-3-oxo-2,3-dihydro-1H-indene-2,1-diylidene))dimalononitrile) binary systems were investigated, along with the ternary PBDB-T:PC$_{71}$BM:IEICO-4F system.

*Solution preparation*: PM6, PBDB-T and IEICO-4F were purchased from Solarmer, PC$_{71}$BM was purchased from the 1Material. Y6 was synthesized according to literature.[29] The solution of binary PBDB-T:IEICO-4F (1:1.5) and ternary PBDB-T:PC$_{71}$BM:IEICO-4F (1:0.7:0.8) for the active layers were prepared both with a concentration of 20mg/mL in total in the mixture of chloroform (CF) and 1-chloronaphthalene (CN) (96 to 4 vol%). PM6 and Y6 were dissolved in CF to a total concentration of 16 mg/mL with a 1 to 1.2 weight ratio and 0.5% CN (v/v, CN/CF) as additive. The solution for the active layers was stirred at least for 5h

*Device fabrication*: Hole-only devices were made with a configuration of glass/ITO/PEDOT:PSS (20nm) /active layer/MoO$_x$ (8nm) /Al (90nm), and the electron-only devices were made with a configuration of glass/ITO/PEIE (2nm) /Active layer/LiF (0.6nm) /Al (90nm). The ITO-coated glasses were boiled in a mixture of deionized water, ammonium hydroxide (25%) and hydrogen peroxide (28%) (5:1:1 vol) at 80°C for 15min for cleaning. PEDOT:PSS (Baytron P VP Al 4083, diluted with deionized water 1:1 vol) was spin-coated onto the ITO glasses at 3000 rpm for 40 s, followed by annealing at 150 °C for 10 min. Polyethylenimine ethoxylated (PEIE), dissolved in isopropyl alcohol (0.05 wt%), was spin-coated onto the ITO glasses at 4000rpm for 30s, followed by annealing at 100°C for 10 minutes. The MoO$_x$/Al and LiF/Al layers were deposited by thermal evaporation through a shadow mask. The active layers were deposited by spin-coating at 2500 rpm for 60 s, followed by annealing at 100°C for 10 minutes.

*Electrical characterization*: Temperature dependent SCLC measurements were performed inside a Janis cryogenic probe station under vacuum (~10$^{-5}$ mbar). Care was taken that jV curves before and after the temperature sweep were essentially identical.

## Results

### Calibration and comparison of the analytical model with kMC

Before using the analytical VRH model to extract parameters from experimental data, we need to identify the values of the parameters that go in the mobility prefactor Eq. 6. To this end, the mobility was studied as a function of concentration and electric field, parametric in $\sigma_{DOS}$, $T$, $a_{NN}$, and $\alpha$, see Fig. 1. The mobilities obtained with the exact kMC model (symbols) were globally fitted with the analytical model (lines) in order to determine the parameters $\lambda_1$, $\lambda_2$, $B$ and $B_c$. We found $B_c \approx 2.7$, in line with previous literature.[10,30] The optimal values of $\lambda_1$ and $\lambda_2$ are 0 and 0.5, respectively, but we should stress that changes in these values only lead to minor, mostly quantitative changes that can be compensated by changing *B*.

Irrespective of the parameter values used, the mobility is constant at low concentrations and then increases in a power law fashion, which is in line previous experimental results [31,32] and consistent with analytical[12] and numerical treatments of the Gaussian disorder model[9]. In the entire parameter space, the difference between the numerical and analytical models is well within a factor two, which is reasonable in view of typical sample-to-sample variations encountered experimentally, with maximum variations occurring at low temperatures. While it is possible to improve the agreement in one panel, doing so would deteriorate the agreement in another and the given parameters reflect a broad global optimum. All in all, the analytical model seems to capture the key traits of the quasi-atomistic simulations in a physically transparent manner with sufficient accuracy to be used in experiment analysis.

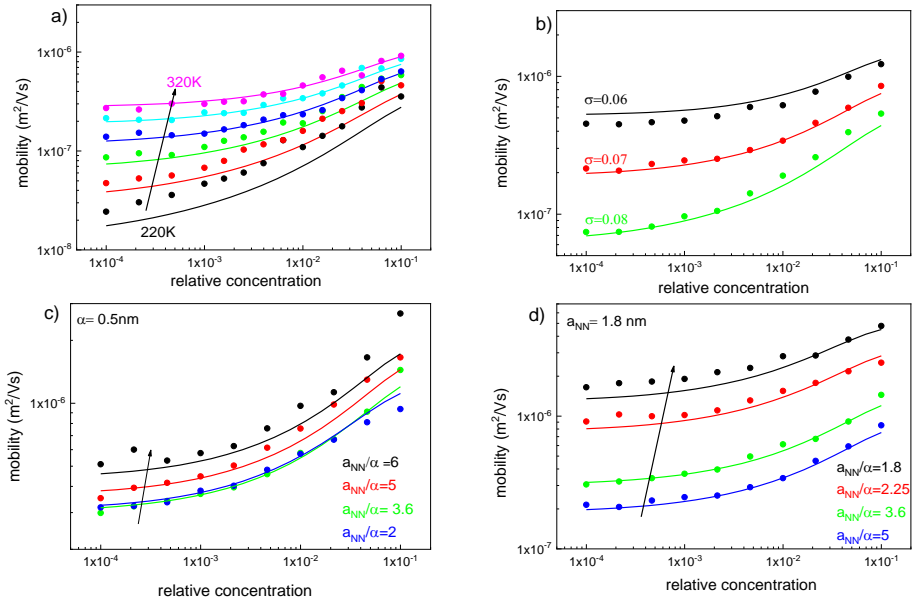

Fig. 1. Analytical model (lines) compared to kMC simulations (symbols) for charge carrier mobility as a function of concentration parametric in temperature (a), disorder (b), mean inter-site distance $a_{NN}$ (c) and localization radius α (d). Parameters in the mobility calculation are $\sigma_{DOS}$=70 meV, T=300 K, $a_{NN}$=1.8 nm and α=0.36 nm unless indicated otherwise. Optimal parameters in the analytical model are $\lambda_1$=0 $\lambda_2$=0.5, $B$=3 and $B_c$=2.7 and are used throughout the text.

Fig. 1c and 1d show the variation of the mobility as a function of $a_{NN}$ and α. In interpreting these figures, the mobility prefactor $v'_0 = v_0 \exp(2a_{NN}/\alpha)$ that is kept constant at $v'_0 = 1\times10^{11}$ s$^{-1}$ should be kept in mind. Specifically, the mobility increase with localization radius in Fig. 1d does not reflect the more or less trivial (exponential) dependence of μ on $\alpha$, but instead reflects the easier long-range percolation due to variable range hopping – in the limit of strict nearest neighbor hopping the plotted mobility would be independent of $\alpha$. Note also that the distinct dependence of the mobility on $a_{NN}$ (panel c) and $\alpha$ stems to a large degree from the non-exponential terms in the mobility, as can be seen from the comparison with calculations for $\lambda_2$=0 in SI Fig. S1.

Shklovskii et al and Nenashev et al [28] argued that the localization length $\alpha$ and not the typical inter-site distance $a_{NN}$ is the decisive length scale that determines the field dependence of the mobility. In the limit of strict nearest neighbor hopping, as considered by Pasveer, this distinction becomes irrelevant[9]. In Fig. 2 we find good agreement between the field dependence of the mobility as obtained by kMC and by the analytical model using the same parameters as in Fig. 1. In view of the above, the apparent dependence of the field dependence of the mobility on $a_{NN}$ in Fig. 2c may appear counter intuitive. However, although the effective temperature Eq. 7 does not depend on $a_{NN}$ or $\sigma_{DOS}$, c.f. panel b, the dependence of μ on the (effective) temperature does depend on these parameters, rationalizing the different high-field trends in Fig. 2.

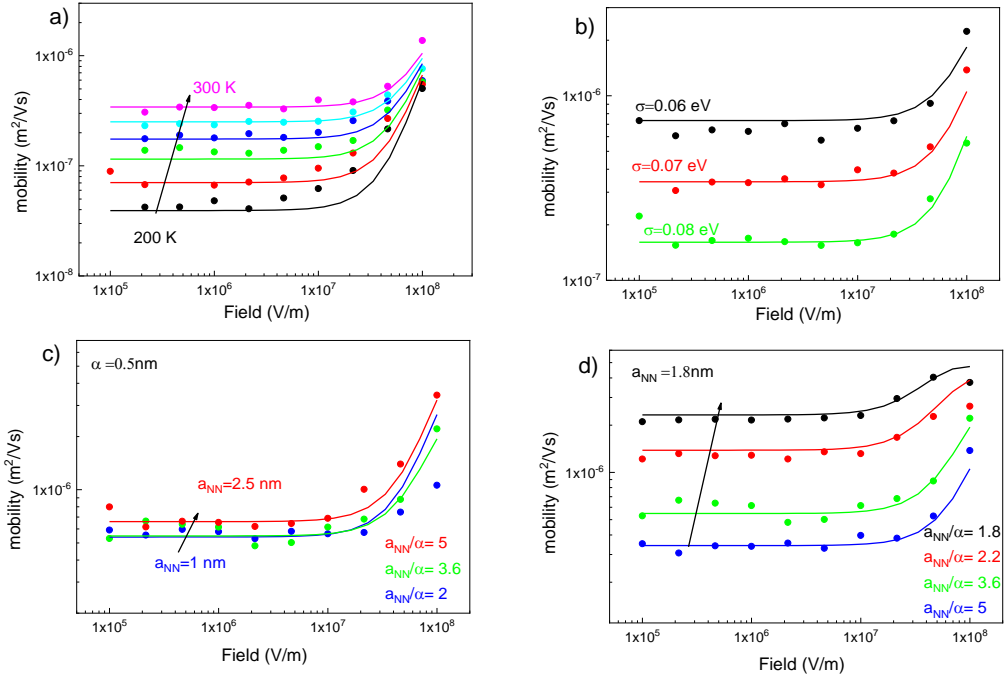

Fig. 2: Analytical model (lines) calibrated with kMC (symbols) for mobility as a function of field with concentration kept at $1\times10^{-2}$, parametric in temperature (a), disorder (b), mean inter-site distance $a_{NN}$ (c) and localization radius α (d). Parameters in the mobility calculation are $\sigma_{DOS}$=70 meV, $T$=300 K, $a_{NN}$=1.8 nm and $\alpha$=0.36 nm unless indicated otherwise.

## Analysis of experimental SCLC data

Fig. 3 shows the analysis of hole- and electron-only devices of the amorphous polymer:fullerene system TQ1:PCBM, using mobilities from both the eGDM and the ET-GDM. For the hole-only devices there is no noticeable difference in fit quality, and concomitantly the extracted disorder values of $\sigma_{HOMO}$ = 85 meV and 78 meV for the eGDM and the ET-GDM, respectively, are rather similar. An overview of all fitted disorder and hopping parameters of all systems investigated is shown in Table S1 in the SI. For the electron-only devices the situation is markedly different, and the eGDM (blue dashes) clearly fails to produce an acceptable fit, whereas the ET-GDM (solid red) does. This is reflected in the dissimilar disorder values of $\sigma_{LUMO}$ = 100 meV and 115 meV for the eGDM and the ET-GDM, respectively. The underlying reason for the failure of the eGDM is apparent from the small value of the ratio $a_{NN}/\alpha$ = 2.0 that indicates that hopping beyond the nearest neighbor is already important around room temperature. As variable range hopping is known to suppress the temperature dependence of the mobility, any model that only accounts for nearest neighbor hopping will (try to) compensate by a low(er) energetic disorder, which also leads to a lower temperature dependence.[9,33] It is in this context important that the fitted ET-GDM value of $\sigma_{LUMO}$ = 115 meV is within experimental uncertainty equal to the value of ~120 meV found from the detailed analysis of dispersive ultrafast charge transport in the same system[5,6].

It is in this context instructive to briefly address the analysis of the data in Fig. 3 using the Murgatroyd model in which an empirical field enhancement of the mobility is integrated into an analytical expression for space charge limited transport; full details are given in the SI section 5, Fig. S3 and Ref.

[21]. Although this procedure leads to excellent fits of both sets of jV data, it yields $\sigma_{HOMO}$ = 87 meV and $\sigma_{LUMO}$ = 82 meV. In other words, it leads to similar or even larger deviations from the actual disorder values than the eGDM, which is not surprising in view of the fact that the temperature dependent zero-field mobility that is extracted from the Murgatroyd fit is analyzed using the classical GDM model.

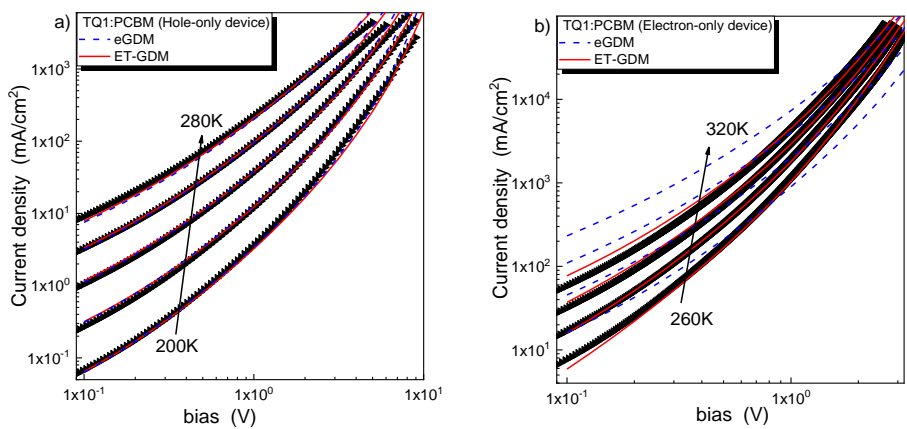

Fig. 3. Experimental jV curves (symbols) and numerical drift-diffusion model fits using mobilities from the eGDM (blue dashed lines) and the ET-GDM (solid red lines) models for TQ1:PCBM hole-only (a) and electron-only (b) devices. Extracted disorder and hopping parameters are listed in Table S1 in section 3 of the SI.

In view of the recent surge in non-fullerene acceptor materials, and the concomitant rise in conversion efficiencies in the OPV field, it is important that the proposed ET-GDM model also provides a consistent description of charge transport in these systems. In addition, reliable disorder values for these novel compounds are rare.

One of the currently best performing binary systems is the PM6:Y6 system, reaching power conversion efficiencies (PCE) in excess of 15% in optimized devices[29]. In our hands, this system has a PCE around 14%, see Table S2 of the SI. Fig. 4 shows the analysis of SCLC data for hole- (panel a) and electron-only (panel b) devices of this blend. For these devices there is no significant difference in the fit quality nor in the extracted disorder and hopping parameters, cf. Table S1. This equivalence is consistent with the $a_{NN}/\alpha$ ratios around 5 that are found from the ET-GDM fit, indicating that around room temperature hopping is predominantly to nearest neighbors.

Intuitively, on basis of the low energy loss in this OPV system one might have expected a relatively small energetic disorder.[29,34] However, we find $\sigma_{HOMO}$ = 89 meV and $\sigma_{LUMO}$ = 71 meV, which, in a quasi-equilibrium view of OPV devices, would correspond to a disorder-induced voltage loss in excess of 0.2 V.[35] Although this is beyond the scope of the present work, we note that the high PCE values observed for this system are therefore unlikely to be related to a suppressed energetic disorder.

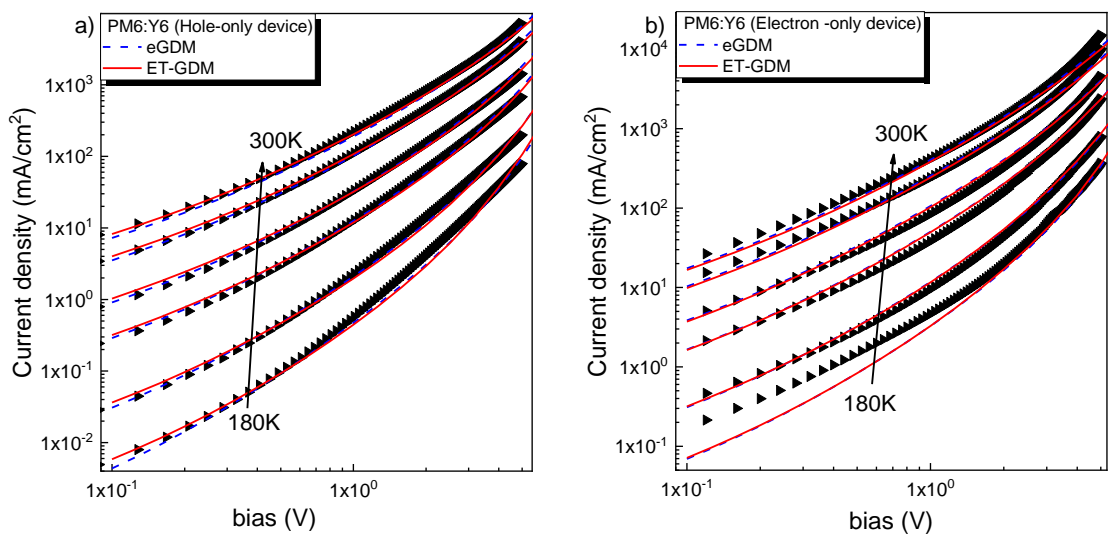

Fig. 4. Experimental jV curves (symbols) and numerical drift-diffusion model fits using mobilities from the eGDM (blue dashed lines) and the ET-GDM (solid red lines) models for PM6:Y6 hole-only (a) and electron-only (b) devices. Extracted disorder and hopping parameters are listed in Table S1 in section 3 of the SI.

Although the underlying mechanisms are incompletely understood, ternary blends of two donors and one acceptor, or vice versa, have proven a viable route to PCE values that exceed those of the corresponding binary compounds.[36,37] In a recent work, some of us argued that the quasi-linear composition dependence of the open circuit voltage, which is crucial for the improved PCE of ternaries, is quantitatively related to the finite width of the density of states of the constituent materials and its variation with composition.[38,39] It is therefore of interest to also test the new mobility model on a recent ternary blend consisting of one donor and two acceptors. Temperature dependent charge transport was examined for the PBDB-T:IEICO-4F:PCBM system, using two different recipes (solvent + additive, see experimental section). In addition, the complementary binary PBDB-T:IEICO-4F systems were investigated.[40] Fig. 5 shows a representative example; the SCLC data for the other combinations are shown in the SI section 5.

As for the example in Fig. 5, virtually all of the single carrier SCLC data for this material system are equally well fitted by either of the two mobility models. This optical similarity notwithstanding, the disorder values obtained by the eGDM can differ markedly from those obtained with the ET-GDM. Deviations between the two models clearly correlate with the value of the $a_{NN}/\alpha$ ratio as shown in the SI, Fig. S3, with deviations exceeding 10 meV occurring around $a_{NN}/\alpha = 3$ and below. Hence, as for the Murgatroyd fit discussed in the context of Fig. 3 above, an 'optically' good fit does not guarantee an appropriate model is used.

A factor that can hamper the accurate parameter extraction from SCLC curves are charge carrier traps that may for instance be induced by nanoscopic water clusters and that show up as a 'hump' in the jV curve that cannot, without further additions, be reproduced by a drift-diffusion model with either of the investigated mobility expressions.[41] Alternatively, morphological dead ends may act as charge traps that suppress the current (increase) at higher fields.[42,43] Although not visible in Fig. 5, the PBDB-T:IEICO-4F system has a clear propensity to form various types of traps, especially for electrons, cf.

Figs. S5-7. Although this does not correlate with an increase in photovoltaic performance (Table S2), we note that the addition of PCBM to the layer deposited from chloroform seems to suppress the traps. A similar effect of PCBM was observed in Ref. [44]. A hand-waiving argument for this behavior would be that PCBM occupies the voids that originally were occupied by water and thereby gives a more compact morphology with a lower concentration of traps.[41,45]

In view of the uncertainties induced by the charge trapping in this system we refrain from a detailed interpretation of specific changes of parameters in response to changes in processing or composition. We do, however, notice that the disorder parameters of this system, as for the other systems we looked at here, are rather unremarkable in the range 50-80 meV, with $\sigma_{LUMO}$ typically being somewhat smaller than $\sigma_{HOMO}$ and substantially smaller than what is typically found for PCBM, here and elsewhere.[5,6,46] We finally note that the addition of PCBM to the binary system does not seem to lead to an appreciable increase in $\sigma_{LUMO}$, which at least in part can be rationalized by the fact that the LUMO level of the IEICO-4F lies below that of PCBM so that electron transport will predominantly take place through the IEICO-4F subphase.[40]

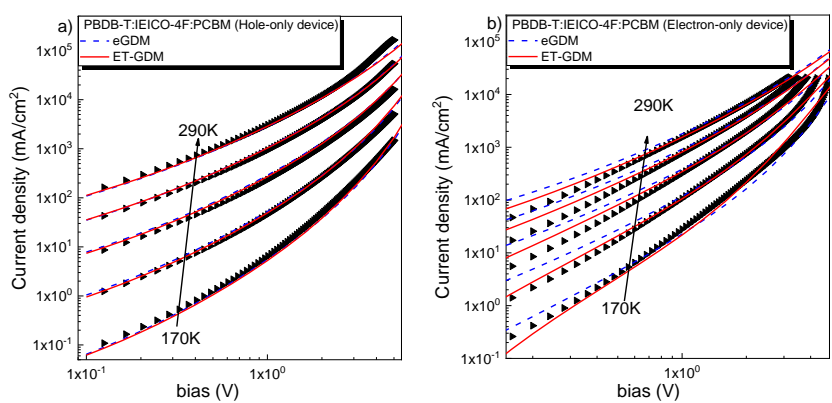

Fig. 5. Experimental jV curves (symbols) and numerical drift-diffusion model fits using mobilities from the eGDM (blue dashed lines) and the ET-GDM (solid red lines) models for PBDB-T:IEICO-4F:PCBM (CF+4% CN) hole-only (a) and electron-only (b) devices. Extracted disorder and hopping parameters are listed in Table S1 in section 3 of the SI.

## Conclusions

We have presented a variable range hopping model for electron and hole transport in energetically disordered organic semiconductors. Whereas the exponential terms in the model are based on previous literature, the algebraic mobility prefactor was heuristically obtained and calibrated by direct comparison between the analytical model and kinetic Monte Carlo simulations over a relevant range of parameters. Good agreement between the two models was obtained. The analytical mobility expressions were subsequently integrated in a freeware drift-diffusion solver that can be used for automated fitting of temperature dependent space charge limited current data.[21,22]

Analysis of experimental current-voltage curves showed that the proposed model can adequately describe electron and hole transport in a wide variety of organic semiconductor blends as used as active layer in typical bulk heterojunction organic solar cells. Specifically, transport in binary and ternary blends based on both fullerene and non-fullerene acceptors can consistently be described. The latter is in contrast to the extended Gaussian disorder model that is based on nearest neighbor hopping on a regular lattice and only gives reliable results around room temperature when the ratio of the typical inter-site distance and the localization radius is larger than ∼3.

Disorder values of all investigated materials fall in the range of 50 – 90 meV, with the noticeable exception of PCBM ($\sigma_{LUMO}$ = 115 meV), without any clear correlation to OPV performance of the particular blend. This work does, however, show that also non-fullerene acceptors that enable PCEs over 14% have significant (several $k_B T$) energetic disorder. Hence the superior performance of these materials is unlikely to be due to suppressed energetic disorder. At the same time, these 'modern' materials should therefore show the same dispersive ultrafast transport and concomitant slow carrier relaxation that was previously found for fullerene acceptors and polymeric donor and acceptor materials.[47] This will be topic of future work.

## Acknowledgements

We gratefully acknowledge funding by Vetenskapsrådet, project 'OPV2.0'. D.S acknowledges funding from the European Union's Horizon 2020 research and innovation program under the Marie Skłodowska-Curie grant agreement No 799477 — HyThermEl. We thank J. Yuan for providing the Y6 material.